\documentclass[aps,prb,twocolumn]{revtex4-1}
\usepackage{physics}
\usepackage{lipsum}
\usepackage{braket}
\usepackage[pdftex]{graphicx,color}
\usepackage{amsfonts,amsmath,amssymb,bm,amsthm}
\usepackage[colorlinks]{hyperref}

\begin{document}
\title{Phase diagram of multi-layer ferromagnet system with dipole-dipole interaction}
\author{Taichi Hinokihara$^{1,2}$}
\email{thinokihara@issp.u-tokyo.ac.jp}
\author{Seiji Miyashita$^{1,2}$}
\email{miyashita@phys.s.u-tokyo.ac.jp}
\affiliation{%
$^1$Department of Physics, Graduate School of Science,
The University of Tokyo, 7-3-1 Hongo, Bunkyo-Ku, Tokyo 113-8656, Japan\\
$^2$Elements Strategy Initiative Center for Magnetic Materials (ESICMM), National Institute for Materials Science, Tsukuba, Ibaraki, Japan\\
}
\begin{abstract}
We investigate various magnetic configurations caused by the dipole--dipole interaction (DDI) in the thin-film magnet with the perpendicular anisotropy under the open boundary conditions.
Two different approaches are simulated: one starts from a random magnetic configuration and decreases temperatures step-wisely; the other starts from the saturated out-of-plane ferromagnetic state to evaluate its metastability.
As typical patterns of magnetic configuration, five typical configurations are found: an out-of-plane ferromagnetic, in-plane ferromagnetic, vortex, multi-domain, and canted multi-domain states.
Notably, the canted multi-domain forms a concentric magnetic-domain-pattern with an in-plane vortex structure, resulting from the open boundary conditions.
Concerning to the coercivity, a comparison of the magnetic configurations in both processes reveals that the out-of-plane ferromagnetic state exhibits metastability in the multi-domain state, while not in the vortex state.
We also confirm that the so-called Neel-cap magnetic-domain-wall structure, which is originally discussed in the in-plane anisotropy system, appears at the multi-domain state.
\end{abstract}
\maketitle
\section{Introduction}\label{sec:introduction}
Various magnetic configurations appearing in magnetic systems with the dipole-dipole interaction (DDI) are attracting considerable attention not only in the fields of science but also in various industrial purposes.
Even in the pure DDI system~\cite{Luttinger1946,DeBell2000,Takayama2007}, structures and sizes dependence of spin alignment have been investigated, e.g., in single-molecular magnets~\cite{DIC2000,Luis2005} and in high-density magnetic storage.
Moreover, the interplay between the short-range interaction and DDI leads to more complex magnetic properties.
Especially, thin-film systems have been studied both theoretically~\cite{Abanov1995, MacIsaac1995, MacIsaac1996, Pighin2010, Pighin2012, Yokota2018, Komatsu2019, Santamaria2000, DeBell2000, Carubelli2008, Whitehead2008} and experimentally~\cite{Hubert1998,Kooy1960,Slonczewski1973,Taniuchi2015, Taniuchi2016},
e.g., concerning the spin reorientation transition between the in-plane ferromagnetic state and the out-of-plane ferromagnetic state~\cite{Santamaria2000, DeBell2000, Carubelli2008, Whitehead2008}.

Most of these theoretical works have been studied in systems with periodic boundary conditions.
Under the conditions, there exist four magnetic configurations: the out-of-plane ferromagnetic state, the in-plane ferromagnetic state, the multi-domain state with stripe pattern, and the canted stripe state which has been recently discovered between the multi-domain and the in-plane ferromagnetic state~\cite{MacIsaac1996,DeBell2000,Komatsu2019,Whitehead2008, Pighin2010, Pighin2012}.

The study of systems under periodic boundary conditions reveals the magnetic configurations in large-size materials.
However, most permanent magnets consist of a large number of grains.
In the theoretical study for magnetic configuration in one grain, it is essential to simulate the system under the open boundary conditions.
The size and shape of the grain affect the magnetic configuration due to the long-range nature of DDI.
This fact makes the magnetic properties different from those under the periodic boundary conditions.

Several magnetic configurations have been pointed out in thin-film systems under the open boundary conditions: out-of-plane ferromagnetic state, in-plane ferromagnetic state, vortex state, and multi-domain state~\cite{Sasaki1997, Matsubara2004, Li2002}.
Especially in the thin-film systems, the vortex structure widely appears in a weak anisotropy region, because this configuration can reduce the stray field.
The vortex state will generate a kind of canted spin state deducing from the canted stripe state in the periodic boundary conditions.
However, less is known about the characteristics of a canted spin state under the open boundary conditions.
In the present paper, we systematically study how magnetic configurations change with different shapes and sizes of systems due to the long-range nature of DDI.
To investigate parameter dependence of characteristic magnetic configurations, we survey the magnetic profiles under the open boundary conditions with different anisotropy $K$, DDIs $D$, the exchange coupling $J$, and the thickness of the system $L_z$.
Here, $J$ denotes the strength of the nearest-neighbor coupling, which corresponds to the stiffness constant $A$ in the continuous spin model.

Moreover, the metastability of the out-of-plane ferromagnetic state is not widely understood,
although it is a recent critical topic of the coercivity of permanent magnet~\cite{Fu1992, Tsukahara2019}.
Microscopic observations of domain structure by XMCD (X-ray magnetic circular dichroism) visualizes that various magnetic grains exhibit the multi-domain structure at demagnetized state~\cite{Taniuchi2015, Taniuchi2016, Suzuki2016, Kotani2018, Billington2018, Suzuki2018}. 
As an example, Nd$_2$Fe$_{14}$B magnet consists of micron order magnetic grains, and most of these grains exhibit the multi-domain structure after the thermal demagnetization process~\cite{Kotani2018}.
However, once the system is magnetized by applying a strong magnetic field, it shows a certain amount of coercivity.
Although the metastability of ferromagnetic state in nanocube systems 
has been extensively studied~\cite{Toga2020, Nishino2020}, investigating the mechanism of coercivity in larger systems showing multi-domain structure is also a subject to be investigated. 

In this paper, by using a Monte Carlo (MC) simulation, we present magnetic configurations in $(K/J, D/J)$ space for various thicknesses in two different approaches:
one starts from a random spin configuration and decreases temperatures step-wisely, which we call ``thermal-quench process''; the other starts from the saturated out-of-plane ferromagnetic state to evaluate its metastability, which we call ``field-quench process''.
We consider the thermal quench process (field quench process) corresponds to the thermal demagnetization process (remanent magnetization process) in experiments.
Firstly, we discuss the stational state at a given temperature by using the thermal quench process.
Secondly, we evaluate the metastability of the out-of-plane ferromagnetic state by comparing magnetic configurations obtained by the two different approaches.
We also discuss the energetical structure of magnetic configurations to evaluate the coercivity in detail.

We find the following three novel properties in the magnetic system under the open boundary conditions.
First, we find a canted multi-domain region in between the vortex state and the multi-domain state.
This magnetic configuration shows both a concentric magnetic-domain pattern along the perpendicular axis and an in-plane vortex structure.
Second, in the multi-domain state appearing at thick systems, e.g., $L_z=15$, the magnetic domain-wall (DW) shows a so-called Neel cap structure, which is mainly discussed in the in-plane anisotropic thin-film systems.
This structure shows a wide Neel-type DW in surface of the, while a narrow Bloch-type DW in bulk.
This magnetic structure takes place to reduce the stray field as schematically studied before~\cite{Hubert1998, Slonczewski1973}. 
Third, the metastability of the ferromagnetic state exists in the multi-domain state.
However, other states, such as the vortex state, show no or too small metastability.

For the present study, an efficient numerical method is desirable to calculate long-range interacting systems.
Simulating the long-range interacting system is one of the challenging problems in computational physics because Monte Carlo simulation naively costs $O(N^2)$ computational time, where $N$ is the number of spins in the system. 
To avoid this difficulty, we adopt the recently developed method called the stochastic cut off (SCO) method.
This method enables us to simulate this system with $O(\beta N\ln N)$ computational time~\cite{Sasaki2008, Endo2015, Hinokihara2018}, where $\beta$ is the inverse temperature.

This paper is organized as follows.
In Sec.~\ref{sec:model}, we introduce the model Hamiltonian and briefly explain the SCO method.
In Sec.~\ref{sec:thermal}, we present the magnetic configurations obtained by the thermal-quench process and discuss typical magnetic configurations appearing in systems with the open boundary conditions.
The DW structure in the multi-domain state is also discussed in this section.
In Sec.~\ref{sec:field}, the magnetic configurations obtained by the field-quench process are given, and metastability of the out-of-plane ferromagnetic state is discussed.
We discuss the size-scalability of the present system in Sec.~\ref{sec:scale}.
In Sec.~\ref{sec:conclusion}, conclusion and discussion are given with brief results for three-dimensional systems.

\section{Model and Method}\label{sec:model}
\subsection{Model}
We investigate the following classical Heisenberg spin model with a simple cubic lattice system:
\begin{align}
\mathcal{H} &= -\sum_{i,j} J_{ij} \bm{s}_i\cdot\bm{s}_j - \sum_{i}K s_{zi}^2
+\sum_{i,j}V\qty(\bm{s}_i,\bm{s}_j)\\
V\qty(\bm{s}_i,\bm{s}_j)&= D \qty(\frac{\bm{s}_i\cdot\bm{s}_j}{r_{ij}^3}-3\frac{\qty(\bm{s}_i\cdot \bm{r}_{ij})\qty( \bm{s}_j\cdot \bm{r}_{ij})}{r_{ij}^5}),
\label{hamiltonian}
\end{align}
where $J$, $K$, and $D$ denote the exchange coupling, the uniaxial anisotropy, and the strength of DDI, respectively;
$\bm{r}_{ij}$ denotes the distance vector between the $i$-th and $j$-th spins.
Here, we set the lattice constant $a$, i.e., the distance between the nearest neighbor spins, to be 1.
We also set the spin length $M_i=|\bm{s}_i|$ to be one.
Throughout this paper, we set the exchange coupling $J$ as unit of the energy.

\subsection{SCO method}
As mentioned in the introduction,
The present model suffers from computational difficulty due to the long-range nature of DDI.
To overcome this difficulty, we adopt the SCO method\cite{Mak2005,Sasaki2008,Hinokihara2018}.
Let us briefly explain this method.

The SCO method introduces the bond-update process before the spin-update process.
This bond-update process adopts a pseudo-interaction $\overline{V}$ replacing the original interaction $V$ with probability $p$, and excludes the interaction with probability $1-p$.
It was found that the detailed balance condition of the original system is held by setting $\overline{V}$ and $p$ as follows:
\begin{align}
\label{eq:scoV}
\overline{V}\qty(\bm{s}_i,\bm{s}_j) &= V\qty(\bm{s}_i,\bm{s}_j) - \frac{1}{\beta} \ln\qty[1-p\qty(\bm{s}_i,\bm{s}_j)],\\
p\qty(\bm{s}_i,\bm{s}_j) &= \exp\qty[\beta\qty(V\qty(\bm{s}_i,\bm{s}_j) -V^\ast)],
\label{eq:scop}
\end{align}
where $V^\ast$ is a constant which equals to (or greater than) the maximum value of $V\qty(\bm{s}_i,\bm{s}_j)$ over all the bonds $\bm{s}_i$ and $\bm{s}_j$.
Thus, the stationary state of the simulation is guaranteed to be the same as equilibrium state of the original model. 

Previous studies have proposed algorithms for efficient bond updating~\cite{Sasaki2008,Hinokihara2018}.
Because the bond update process rarely picks up long-distant weak bonds, according to Eq.~\eqref{eq:scop}, a drastic reduction of overall computational time is realized.
As an example, for three-dimensional DDI system, one MC step can be computed in $\mathcal{O}\qty(\beta N \ln N )$, where $N$ denotes the number of spins in the system.

\begin{figure*}[t!]
	\centering
	\begin{tabular}{c}
		\begin{minipage}{0.24\hsize}
			\centering
			\textbf{$L_z=1$}
			\includegraphics[keepaspectratio,width=\hsize]{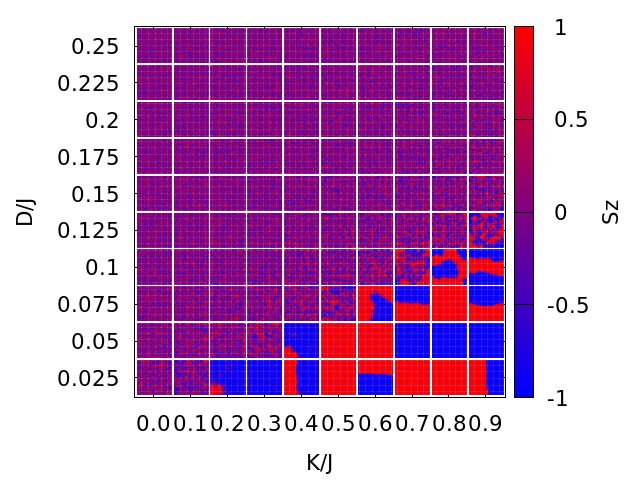}
		\end{minipage}
		\begin{minipage}{0.24\hsize}
			\centering
			\textbf{$L_z=5$}
			\includegraphics[keepaspectratio,width=\hsize]{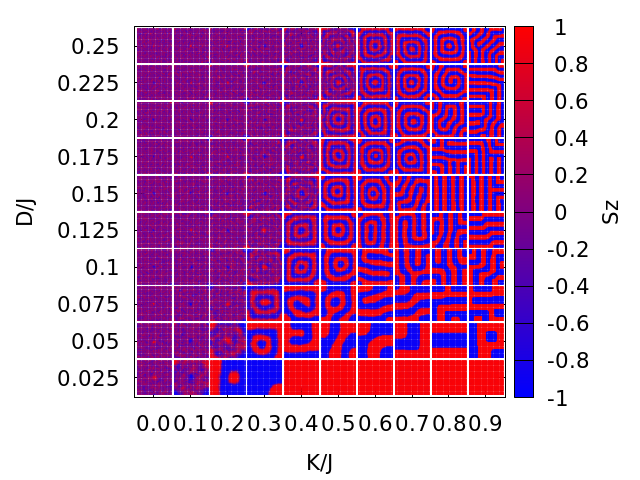}
		\end{minipage}
		\begin{minipage}{0.24\hsize}
			\centering
			\textbf{$L_z=10$}
			\includegraphics[keepaspectratio,width=\hsize]{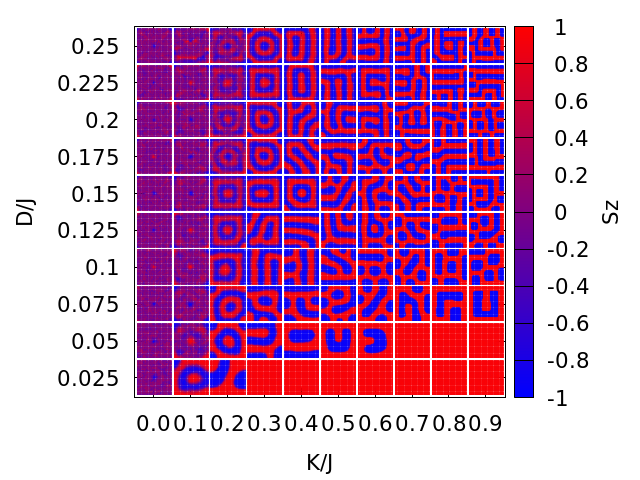}
		\end{minipage}
		\begin{minipage}{0.24\hsize}
			\centering
			\textbf{$L_z=15$}
			\includegraphics[keepaspectratio,width=\hsize]{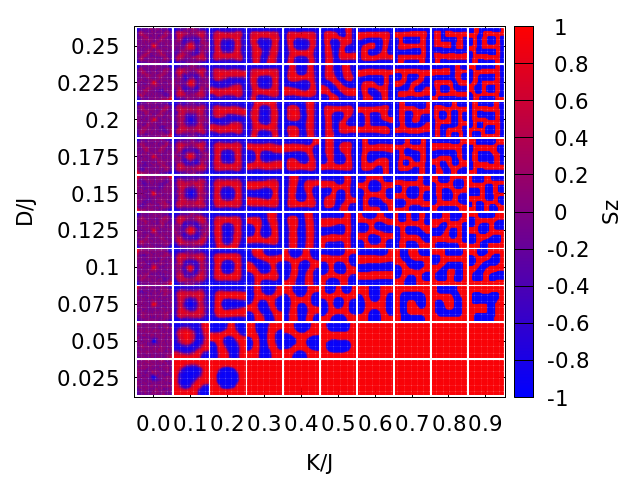}
		\end{minipage}
		\\
		\begin{minipage}{0.24\hsize}
			\centering
			\includegraphics[keepaspectratio,width=\hsize]{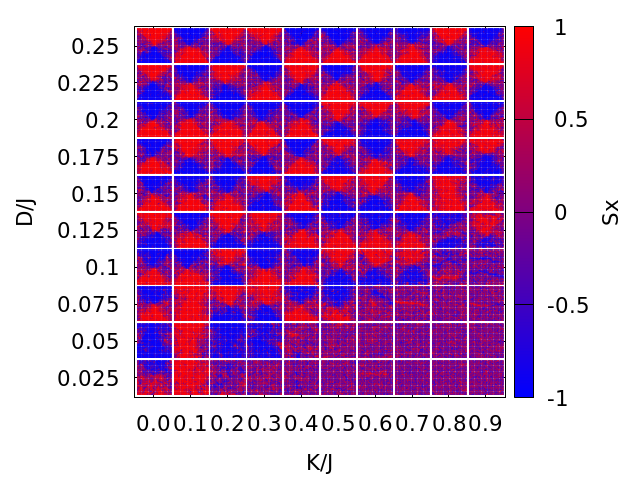}
		\end{minipage}
		\begin{minipage}{0.24\hsize}
			\centering
			\includegraphics[keepaspectratio,width=\hsize]{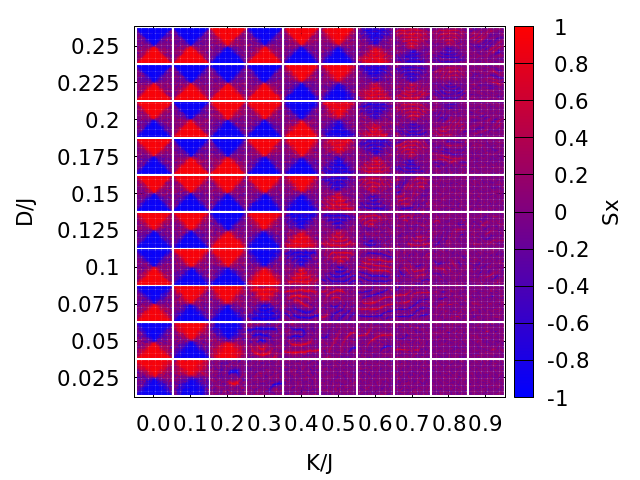}
		\end{minipage}
		\begin{minipage}{0.24\hsize}
			\centering
			\includegraphics[keepaspectratio,width=\hsize]{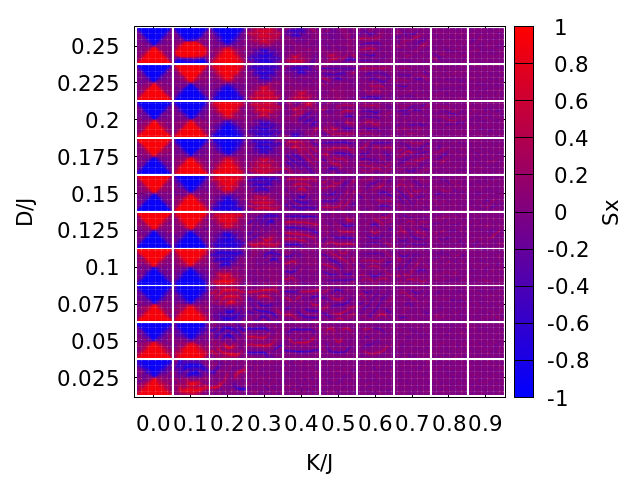}
		\end{minipage}
		\begin{minipage}{0.24\hsize}
			\centering
			\includegraphics[keepaspectratio,width=\hsize]{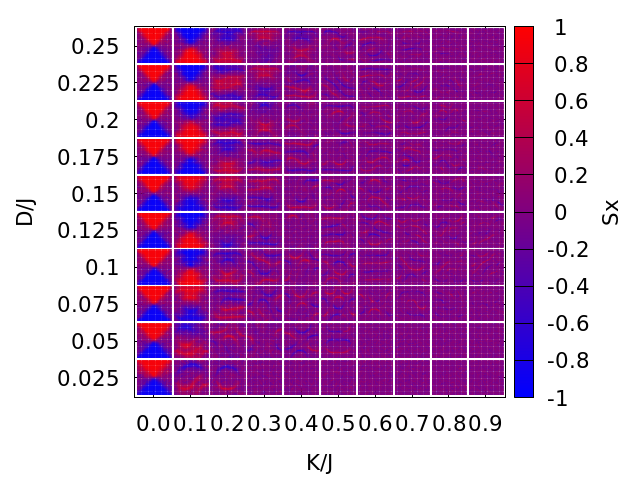}
		\end{minipage}
	\end{tabular}
	\caption{Magnetic structures under the cooling process with different thicknesses, anisotropies, and DDIs.
	Out-of-plane component (top panels) and the in-plane horizontal component (bottom panel) are exhibited. The temperature is $0.3J$  }
	\label{fig:eqstat}
\end{figure*}
\section{Magnetization configurations in thermal quench process} \label{sec:thermal}

Let us first present the magnetic configurations, which appear in the thermal quench process, in the parameter space $(K/J,D/J)$.
Hereafter, we adopt the temperature as $T=0.3J$, which is lower than the critical temperature $T_c$ for both two and three dimensional systems.
To produce the magnetic configurations in the demagnetization process, we simulate the following thermal quench process:
first we perform 50,000~MCS at $T=1.5J$, and then simulate 10,000~MCS at temperatures $T=0.8J$, $0.5J$, and $0.4J$, and finally 50,000~MCS at $T=0.3J$.
We confirmed that details of this process do not affect significantly.

In Fig.~\ref{fig:eqstat}, the upper (lower) panel shows the configuration of the $z$ ($x$) component of $64\times 64\times L_z$ systems as a function of anisotropy ($K/J$) and DDI ($D/J$), in a style of a phase diagram. 
Here, the values of $\langle S_z\rangle$ and $\langle S_x\rangle$ is the averaged magnetizations over the layers.
It should be noted that some systems do not necessarily reach an equilibrium state, especially near the boundaries of two different spin configurations.
However, qualitative information of the phase diagram is well observed.

\begin{figure*}[t!]
\centering
\includegraphics[keepaspectratio,width=\hsize]{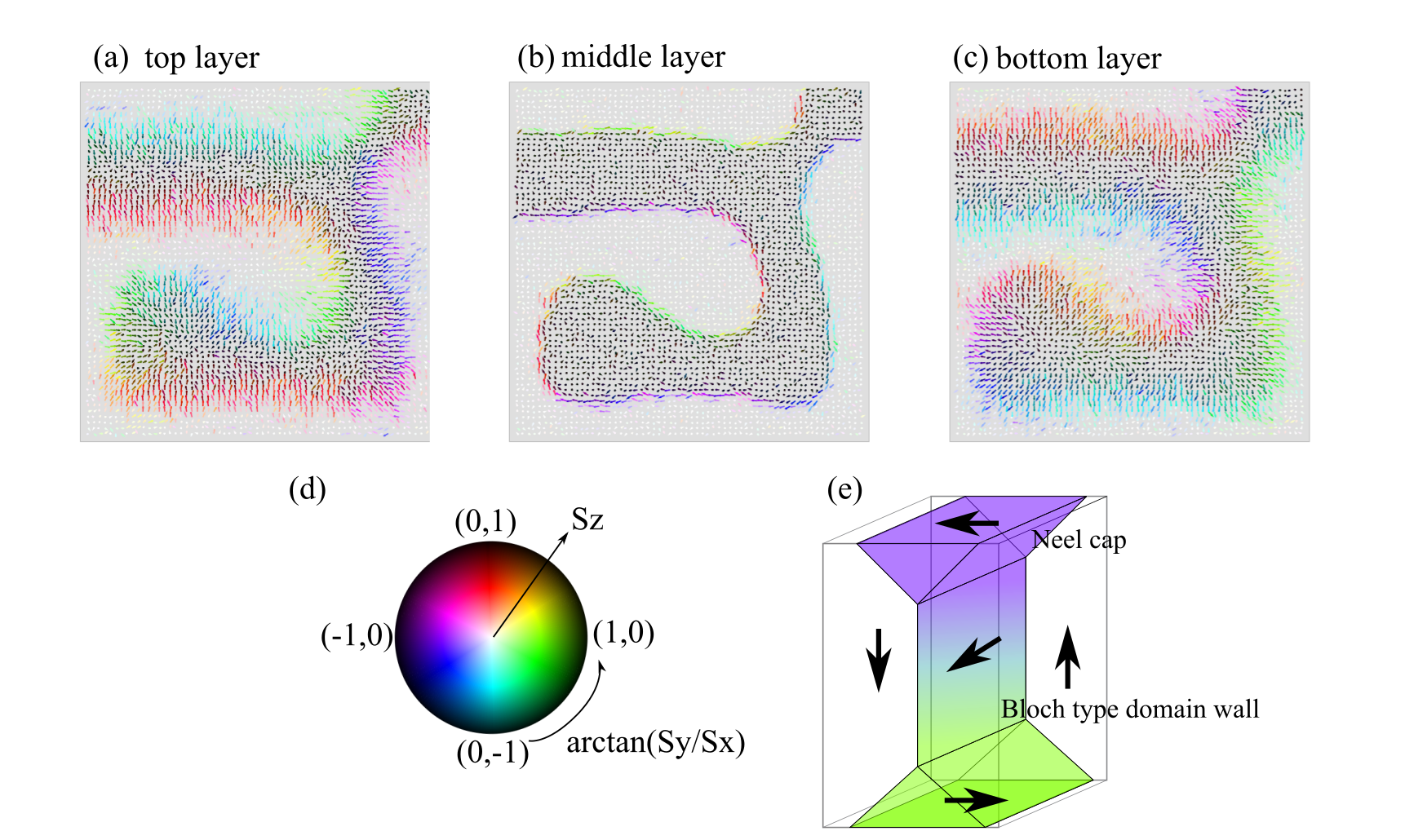}
\caption{Magnetization of the top layer (a), middle layer (b), and the bottom layer (c) of the $64\times64\times15$ system with $D=0.175J$ and $K=0.4J$.
	The magnetization direction is depicted following the color map (d). The schematic  picture of the domain wall is shown in (e).
}
\label{fig:dw}
\end{figure*}

\subsection{magnetic configurations}
In Fig.~\ref{fig:eqstat}, we find five qualitatively distinct magnetic configurations:
\begin{description}
\item{\bf{Region (i): out-of-plane ferromagnetic state}}\mbox{}\\
All the spins are oriented in the easy-axis direction. 
This state is stable in a region of weak DDI and strong anisotropy $K$. 

\item{\bf{Region (ii): in-plane ferromagnetic state}}\mbox{}\\
All the spins are uniformly oriented to the in-plane axis.
This state appears only in $L_z=1$ systems with weak DDI and weak anisotropy, such as the region of $D<0.05J$ and $K<0.2J$, 
and rapidly disappears in the current parameter range as the system thickens.

\item{\bf{Region (iii): vortex state}}\mbox{}\\
Most of the spins are oriented in the plane and form a vortex structure.
At the vortex center, spins tend to be oriented perpendicular to the plane.
This state shows not a circular vortex but an 'X'-like pattern reflecting the shape of the square disk system.
The vortex state is stable in a region with weak anisotropy and strong DDI.
This state is widely stable in the thin system but rapidly shrinks to a weak DDI region as $L_z$ increases.

\end{description}

In Fig.~\ref{fig:eqstat}, above the region (i), magnetic configurations show multi domains with opposite out-of-plane magnetizations.
The domain patterns form either a concentric, a stripe, or a maze-like patterns.
We divide this multi-domain region into two parts, depending on whether the vortex structure appears in the $xy$ components.

\begin{description}
\item{\bf{Region (iv): multi-domain state}}\mbox{}\\
In the strong anisotropy region, a complex $z$-component order with maze-like or stripe patterns appears.
Most of the spins are oriented along the easy axis ($z$-direction), and thus no typical in-plane magnetic structures except for nearby the DW.
In the case of $L_z=1$, the multi-domain state does not appear in the present parameter range.
We confirmed that it appears in much stronger DDI and anisotropy region (not shown).
The interval of magnetic domains, i.e., the width of stripes, becomes narrower as the DDI increases, while it becomes wider as $L_z$ increases.

\item{\bf{Region (v): canted multi-domain state}}\mbox{}\\
Between the regions (iii) and (iv), e.g., a configuration of $K=0.3J$, $D=0.075J$, and $L_z=5$, 
we find a concentric magnetic-domain pattern in $z$-components.
Most of these concentric magnetic-domain patterns spontaneously show a vortex structure in the $xy$-components.
Namely, the spins are canted from the perpendicular axis to the surface.
We consider this state appears by the same mechanism as the canted stripe state, which appears in the system under the periodic boundary conditions~\cite{Pighin2010,Pighin2012,Whitehead2008}.
Namely, the canted stripe state apperas by the spin reorientation transition between the stripe state and the in-plane state.
However, reflecting the nature of the open boundary conditions, the magnetic-domain pattern is different from the canted stripe state.

\end{description}

\begin{figure*}[t!]
\centering
\begin{tabular}{c}
\begin{minipage}{0.24\hsize}
\centering
\textbf{$L_z=1$}
\includegraphics[keepaspectratio,width=\hsize]{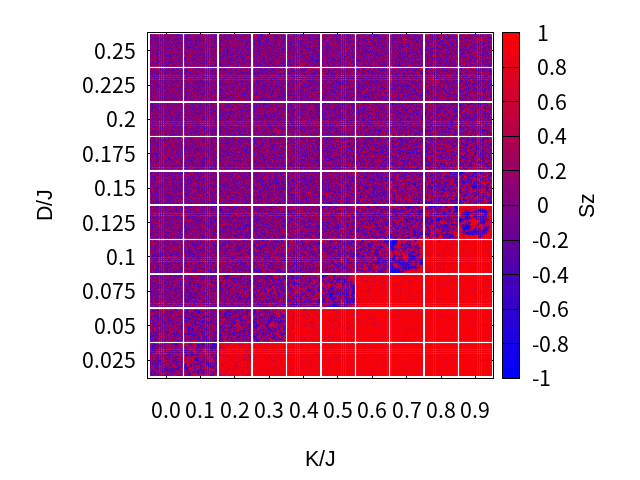}
\end{minipage}
\begin{minipage}{0.24\hsize}
\centering
\textbf{$L_z=5$}
\includegraphics[keepaspectratio,width=\hsize]{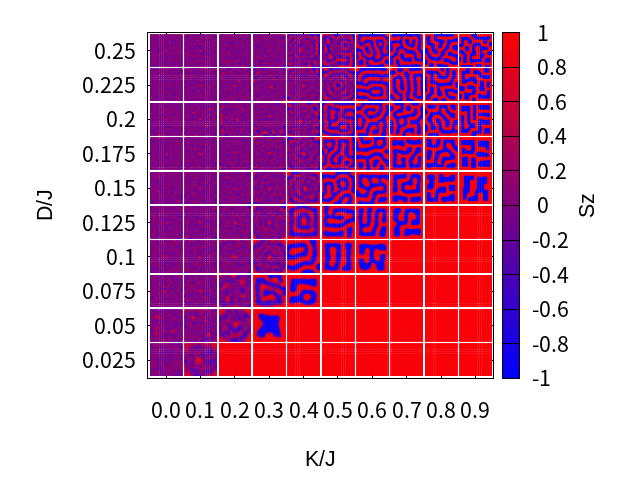}
\end{minipage}
\begin{minipage}{0.24\hsize}
\centering
\textbf{$L_z=10$}
\includegraphics[keepaspectratio,width=\hsize]{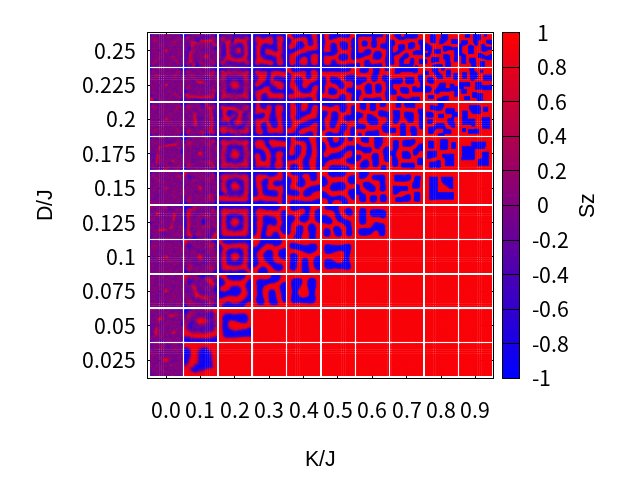}
\end{minipage}
\begin{minipage}{0.24\hsize}
\centering
\textbf{$L_z=15$}
\includegraphics[keepaspectratio,width=\hsize]{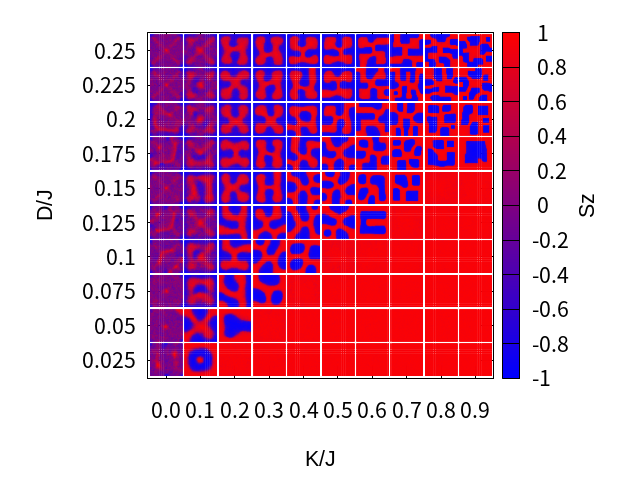}
\end{minipage}
\\
\begin{minipage}{0.24\hsize}
\centering
\includegraphics[keepaspectratio,width=\hsize]{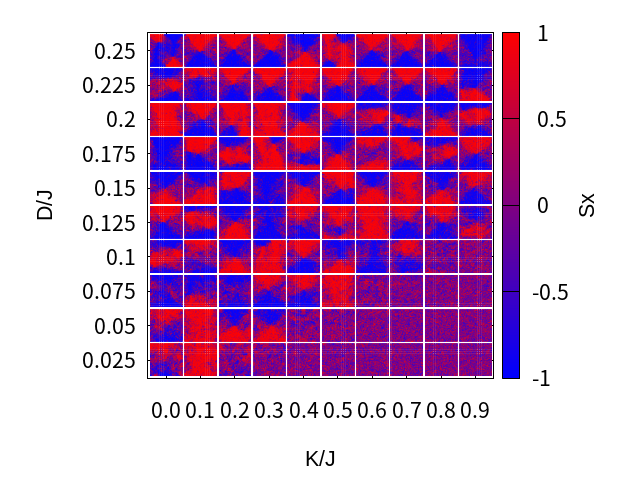}
\end{minipage}
\begin{minipage}{0.24\hsize}
\centering
\includegraphics[keepaspectratio,width=\hsize]{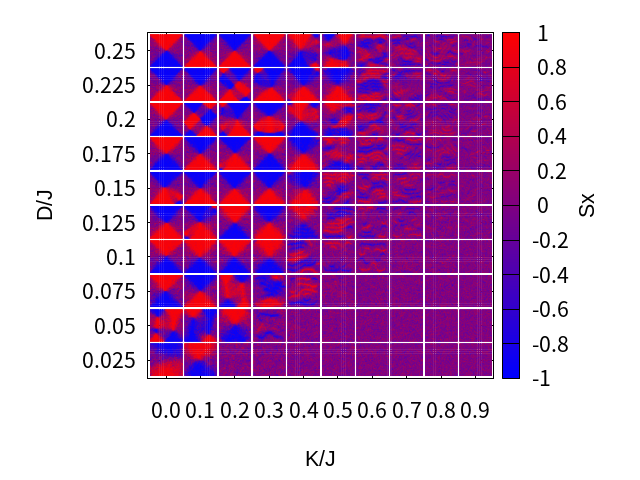}
\end{minipage}
\begin{minipage}{0.24\hsize}
\centering
\includegraphics[keepaspectratio,width=\hsize]{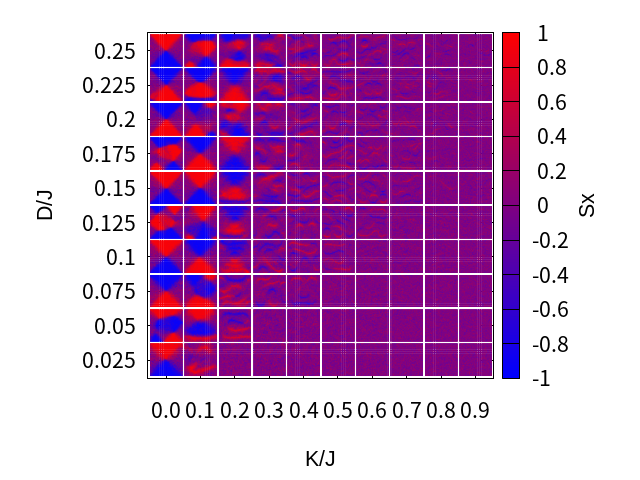}
\end{minipage}
\begin{minipage}{0.24\hsize}
\centering
\includegraphics[keepaspectratio,width=\hsize]{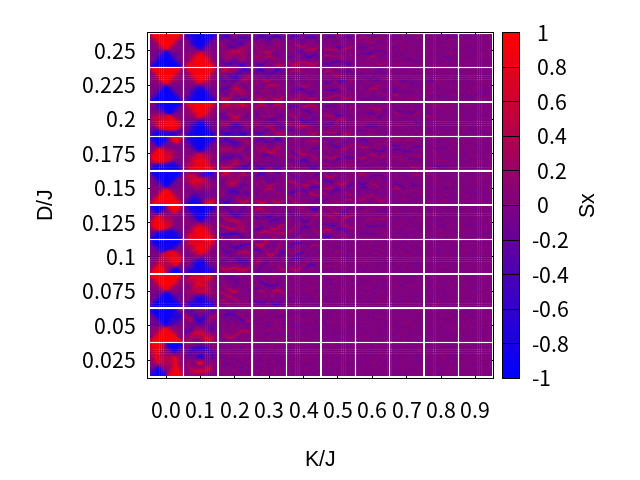}
\end{minipage}
\end{tabular}
\caption{Magnetic structures under the field quench process from the out-of-plane ferromagnetic state with different thicknesses, anisotropies, and DDIs.
Out-of-plane component (top panels) and the in-plane horizontal component (bottom panel) are exhibited. The temperature is $0.3J$}
\label{fig:ferrostart}
\end{figure*}

\subsection{Parameter dependence of magnetic configurations}

We pointed out five typical magnetic configurations above.
Here, let us discuss the dependence of the borders between these configurations.

\subsubsection{Border between (i) and (ii)}
\label{subsub}
The border between the out-of-plane ferromagnetic state (i) and the in-plane ferromagnetic state (ii) is determined by the competition between the anisotropy energy and the DDI energy.
The anisotropy energy lets the system to be the out-of-plane ferromagnetic state, while the in-plane ferromagnetic state is favorable for the DDI energy. 
The total energy per spin for the out-of-plane ferromagnetic state $E_{out}$ and the for the in-plane ferromagnetic state $E_{in}$ can be estimated as 
\begin{align}
	\label{eqout}
	E_{out} &\sim -\frac{\overline{z}}{2}J -K + C_{out}(L_z) D, \\
	\label{eqin}
	E_{in}  &\sim -\frac{\overline{z}}{2}J + C_{in}(L_z)D,
\end{align}
where $\overline{z}$ denotes the average of the number of nearest neighbor spins.
In thin-film systems, $z$ is approximately 4, while $\overline{z}$ approaches to 6 as the system thickens.
$C_{in}$ and $C_{out}$ denote the DDI energies of each magnetic configuration.
These values depend on the size and shape of systems, i.e., the thickness of the system $L_z$ (see Appendix for the thickness dependence of $C_{in}$ and $C_{out}$).

According to Eqs.~\eqref{eqout} and~\eqref{eqin}, these two states linearly changes in the phase space $(K/J,D/J)$ as 
\begin{align}
K=\frac{C_{out}(L_z)-C_{in}(L_z)}{D}.
\label{border1-2}
\end{align}
In Fig.~\ref{fig:eqstat}, we find the in-plane ferromagnetic state only in the system of $L_z=1$ with small $K/J$,
and we cannot identify the border given by Eq.~\eqref{border1-2} because of the vortex state being stable in a broad parameter region instead of the in-plane ferromagnetic state.
We will discuss the border between the vortex state and the others in the following sections.

\subsubsection{Border between (ii) and (iii)}
The total energy per spin for the vortex state $E_{vortex}$ can be estimated as 
\begin{align}
	\label{eqvor}
	E_{vortex} &\sim -\frac{\overline{z}}{2}J +\Delta J + C_{vortex}(L_z) D, \\
\end{align}
where $\Delta$ denotes the loss of the exchange coupling energy due to forming the vortex structure, and $C_{vortex}$ denotes the DDI energy of this state.
First, we mention that $\Delta$ does not depend on the thickness of the system as far as all the spins along the thickness axis are parallel in the vortex state.
In the present system, i.e., $64\times64\times L_z$ system, $\Delta$ is evaluated as $0.0037J$ (we evaluate $\Delta$ by using the spin configurations determined in Appendix).
On the other hand, the DDI energy difference between the in-plane ferromagnetic state and the vortex state is of the order of $0.1D$ for the case of $L_z=1$ (see Fig.~\ref{DDengLz} in Appendix).
Thus, in $L_z=1$ system, these energies, $E_{in}$ and $E_{vortex}$, are same at $D\sim0.035J$.
This result is consistent with the result in Fig.~\ref{fig:eqstat}. 

In thin-film systems, the energies, $E_{in}$ and $E_{vortex}$, have very close values with each other comparing to the out-of-plane ferromagnetic state.
Thus, as the system thickens, the in-plane ferromagnetic state easily changes to the vortex state, as indicated in Fig.~\ref{DDengLz} (Appendix).

\subsubsection{Border between (i) and  (iv)}
In the region of large $K/J$, the out-of-plane ferromagnetic state (i) changes to the multi-domain state (iv) as $D/J$ increases.
The border between them is given by the competition between the energy costs of DW formation and the demagnetization effect due to DDI. 
A naive estimation of costs would be $\sqrt{KJ}$ for the former is of order per the length of DW, and the latter is proportional to $D$.
Thus, for a given size of the system, the border is roughly given by $D\propto \sqrt{KJ}$.
However, precise evaluation of the domain wall energy is difficult because it forms the Neel cap structure in multi-layered systems due to DDI.

In multi-layered systems, the magnetic structure of the domain wall is modified due to the DDI~\cite{Hubert1998,Slonczewski1973,Lu2005}.
Figures \ref{fig:dw} shows the layer dependence of the domain wall structure and its schematic picture for the case of $64\times64\times15$ with $D=0.175J$ and $K=0.4J$.
In the vicinity of the system surface, the domain wall tends to be the Neel type to reduce the stray field known as the Neel cap structure.
These Neel cap structures appear on both top and bottom surfaces in which magnetic moments face to opposite directions to reduce the stray field.
In the middle of the layers, to continuously connect these Neel caps, the domain wall type changes to the Bloch type.
Besides, owing to the DDI effect, the width of the Bloch type domain wall becomes narrower, and thus its formation energy cannot simply estimate as $\sqrt{KJ}$.
Particularly in the strong $K$ region, the domain wall width in bulk becomes a lattice constant, which is called narrow domain-wall.
Then, the formation energy of the narrow domain-wall becomes insensitive to $K$~\cite{Barbara1978,Sasmita2016}.
Consequently, the border (between the out-of-plane ferromagnetic state and the multi-domain state) is also insensitive to $K$.
Since such the domain wall structure requires a certain amount of thickness, the border should be insensitive to $K$ in thick systems.
This behavior is consistent with our results in Fig.~\ref{fig:eqstat}.

\subsubsection{Border between (iii), (iv), and (v)}

In the region between the vortex state (iii) and multi-domain state (iv), we find that the canted multi-domain state (v) appears.
Since we focus on finite-size systems, it is difficult to distinguish whether these borders are a phase transition or a crossover.
In this paper, we distinguish the states (iii) and (v) whether the magnetic domain wall pattern appears, while the states (iv) and (v) whether the vortex order remains.

In the canted multi-domain state, our result indicates that the concentric magnetic domain pattern is stable.
This magnetic domain pattern is far different from the multi-domain state (iv), in which a stripe pattern is considered to be stable.
Indeed, the stripe pattern actually appears at a large $K$ region in the present result.

On the other hand, a maze-like pattern also appears in the multi-domain state with a small $K$ region.
The maze-like pattern may be either an metastable or stable state by itself.
In the former case, due to a large number of metastable magnetic patterns in the multi-domain state,
the system cannot reach to a stable stripe pattern and freeze in a maze-like pattern.
In the latter case, the maze-like pattern is an intrinsic pattern between the concentric and stripe magnetic domain patterns.
In order to clarify whether the maze-like pattern is stable or metastable, further study will be required to clarify this point.

In any case, due to the various magnetic domain wall pattern in the multi-domain state, it is difficult to evaluate its total energy.
However, most of the spins align to the in-plane axis in the vortex state, while they align to the perpendicular axis in the multi-domain state.
Thus, how this border behaves will be roughly understood from the border between the in-plane ferromagnetic state and the out-of-plane ferromagnetic state, which is discussed in Sec.~\ref{subsub}.
Namely, as the system thickens, the spins favor aligning to the perpendicular axis due to the reduction of the demagnetization field.
Therefore, the vortex state changes to the multi-domain state as the system thickens.

\section{Metastability}\label{sec:field}

Next, we show the magnetic structure obtained by the field-quench process from the saturated out-of-plane ferromagnetic state (see Figs.~\ref{fig:ferrostart}).
Here, we set the same parameters as Fig.~\ref{fig:eqstat}, and perform 50,000~MCS after the change of the magnetic field to zero.
Comparing to Figs.~\ref{fig:eqstat} and~\ref{fig:ferrostart}, we find that the out-of-plane ferromagnetic state metastably retains in the multi-domain state.
This metastable region expands as the anisotropy increases or as the system thickens.
On the other hand, the out-of-plane ferromagnetic state does not remain in the vortex state, indicating that the coercivity in the vortex state is zero or too small to observe in the present simulation.

This behavior is consistent with experimental results, i.e., large magnetic grains exhibit
multi-domain state in the thermally demagnetized phase which corresponds to the thermal-quench process, while most of them have a uniform magnetization in the field sweep process from a saturated state, i.e.,
the field-quench process~\cite{Kotani2018}.
The energetical structure of this metastability will be given in the next subsection.

Unlike the thermal-quench process, the concentric multi-domain patterns are not robustly observed in the field-quench process.
For example, the magnetic configuration at $L_z=5$ with $K=0.7J$ and $D=0.225J$, the concentric magnetic pattern appears in the thermal-quench process, while the complex maze structure appears in the field-quench process.
This difference should be attributed to the difference of the relaxation processes.
In the thermal-quench process, the order gradually develops and forms an energetically favorite pattern, while in the field-quench process the ferromagnetic state is destroyed randomly 
at each position of the lattice by the demagnetization field due to DDI. 

Let us study how the metastable out-of-plane ferromagnetic state breaks down.
We find that the magnetization reversal process starts from inside the plane of the system, 
but not from the corners of the system as has been found in small systems~\cite{Toga2020,Nishino2020} where DDI is not relevant. 
For example, $K=0.4J$, $D=0.075J$, and $L_z=10$,
the magnetization at the edges still remains in the up direction, while the magnetization inside the system is already inverted to down direction.
This magnetic configuration clearly indicates that the magnetization reversal process starts not from the corners but from inside the plane.
The similar reversal process is also found in the case of $L_z=64$ with $K=0.5J$ and $D=0.15J$.
We will show an example of configuration just after the collapse of the ferromagnetic state in a process with increasing $D/J$ in Fig.~\ref{fig:nucleationK7}.

\subsection{Energetical study on the metastability}

\begin{figure}[t!] \centering
\includegraphics[keepaspectratio,width=0.75\hsize]{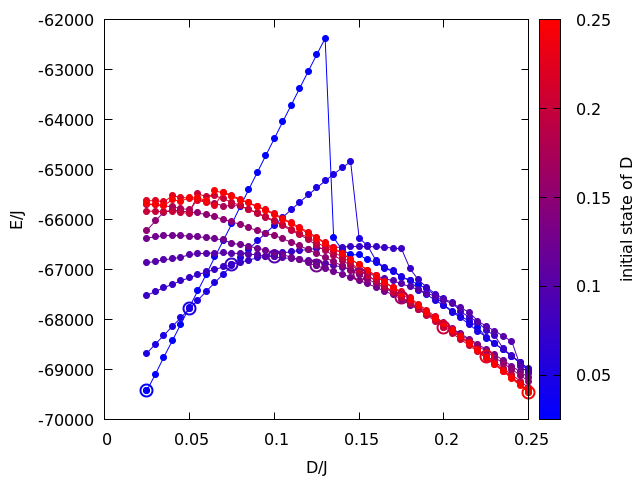}
\caption{Changes of energies of configuration starting from those in Fig.~\ref{fig:eqstat} of various values of $D$ at $K=0.7J$ as a function of $D$ under a sweep of the value of $D$.
Jumps indicates collapses of the original pattern.}
\label{fig:metastle-energy}
\end{figure}

\begin{figure}[h]
\centering
\includegraphics[keepaspectratio,width=0.75\hsize]{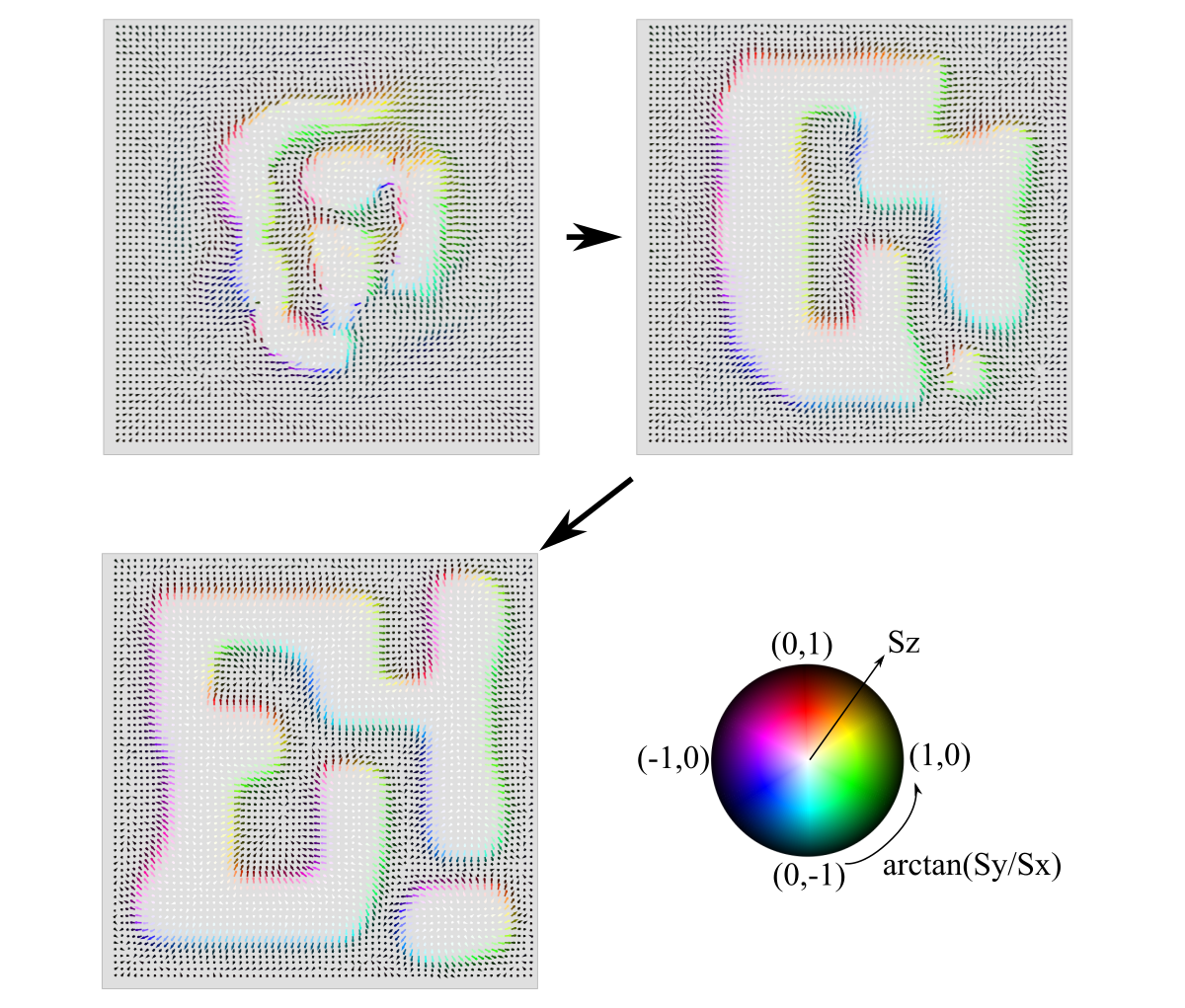}
\caption{Nucleation pattern just after the collapse of the out-of-plane ferromagnetic state with $K=0.7J$ by sweeping the DDI to $D=0.13JD$.}
\label{fig:nucleationK7}
\end{figure}

In order to study the metastability, we study how each characteristic configuration maintains with the variation of the value $D$. 
For this purpose, we performed zero-temperature simulation starting from each configuration given in Fig.~\ref{fig:eqstat} for a given set of $K$ and $D$.
We study how the state changes with varing $D$.
When the value of $D$ reaches to a certain value, the original state abruptly changes. 
The results for $K=0.7J$ are depicted in Fig.~\ref{fig:metastle-energy}.
For each value of $D$, the
energy of the initial configuration given in Fig.~\ref{fig:eqstat} is marked by a point enclosed by a circle.
For example, the energy of the out-of-plane ferromagnetic state which was given in Fig.~\ref{fig:eqstat} at $K=0.7J$ and $D=0.0025J$ is given by blue point enclosed by a circle at most left with the lowest energy. 
We find that the out-of-plane ferromagnetic state survives until $D=0.13J$, and then the configuration collapses to multi-domain state.
Figure~\ref{fig:metastle-energy} indicates that the out-of-plane ferromagnetic state metastably retains even at large values of $D$ where the thermal-quench states are multi-domain state.
This mechanism may give coercivity in rather large grains.
In Fig.~\ref{fig:nucleationK7}, we depict the configuration just after the collapse ($D=0.135J$), where we find the magnetization reversal begins at four points in the plane which is a significant contrast to the case of nanoscale systems where the nucleation begins from corners~\cite{Toga2020,Nishino2020}.

Next, we look at the stability of the multi-domain state at $D=0.25J$.
In Fig.~\ref{fig:metastle-energy} the multi-domain state survives until $D=0.0025J$, although the energy is much higher than others.
This fact indicates that the multi-domain states are deeply metastable.

\section{Size Scalability}\label{sec:scale}

\begin{figure}[t!]
\centering
\begin{tabular}{c}
\begin{minipage}{0.5\hsize}
\centering
\includegraphics[keepaspectratio,width=\hsize]{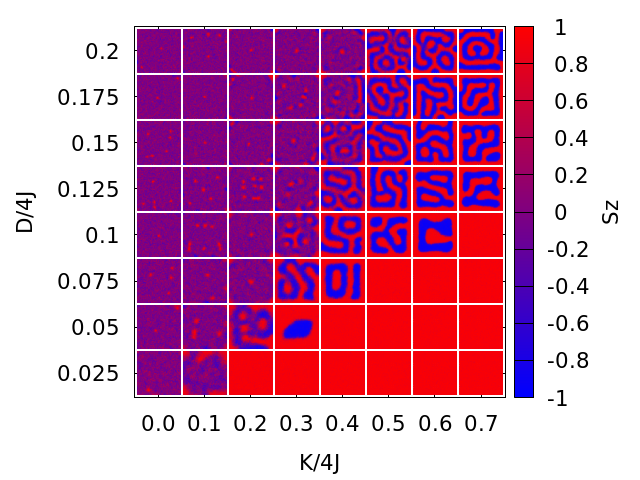}
\end{minipage}
\begin{minipage}{0.5\hsize}
\centering
\includegraphics[keepaspectratio,width=\hsize]{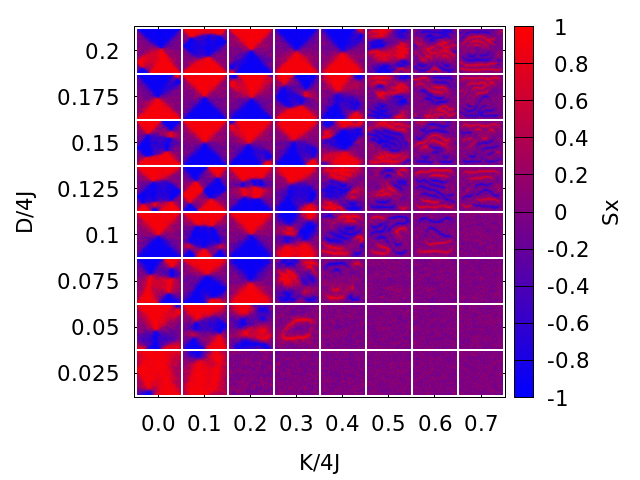}
\end{minipage}
\end{tabular}
\caption{Magnetic structures of the $128\times128\times10$ system under the quench process from the out-of-plane ferro magnetic state with different anisotropies and DDIs.
Out-of-plane component (left panel) and the in-plane horizontal component (right panel) are exhibited. The temperature is $0.3J$ }
\label{fig:128}
\end{figure}

Thus far, we focused on the $64\times64\times L_z$ system.
Because of the peculiar long-range nature of DDI, size dependence is an important issue.
In this section, we study how the magnetic state changes with different sizes keeping the same sample shape (aspect ratio).
The sample-size scaling of the parameters on the lattice constant $a$ in the continuum spin model without thermal fluctuations is well known;
the stiffness constant is proportional to $a$, while the anisotropy energy and the DDI are proportional to $a^3$. 
Namely, when we change the mesh size to be $b$ times larger,
the parameters of this system change to $Kb^3$, $Db^3$, and $Jb$. 
Thus, in the parameter space, $(K/J, D/J)$, should be scaled by 
$(Kb^3/(Ja), Db^3/(aJ))=(Kb^2/J, Db^2/J)$.

In the finite temperature simulations, however, such scaling relation is not ensured.
Thus, we examine to what extent the scaling relation retains at $T=0.3J$.
Figure~\ref{fig:128} shows the magnetic configurations of the $128\times128\times 10$ systems in the quenching process at $T=0.3J$ for the values of $K/J$ and $D/J$. 
We find good agreement with that in Fig.~\ref{fig:ferrostart}.
There the axes of the figure are scaled according to the scaling with $b=2$.
Thus, we find that the size scaling of the micromagnetic model is roughly satisfied.
At $T=0.3J$, the total magnetization of the out-of-plane ferromagnetic state is nearly saturated, i.e., 80\% of the fully magnetic. 
Thus the magnetization does not change drastically before and after the scaling, which causes good agreement.
At higher temperatures, however, the scaling relation must be modified.
Te change of the total magnetization in a unit cell reduces at high temperatures, which also causes renormalization of parameters.
We will study such the scaling-relation in the finite-temperature simulations in the future.

\begin{figure}[t]
\centering
\begin{tabular}{c}
\begin{minipage}{0.45\hsize}
\centering
\includegraphics[keepaspectratio,width=\hsize]{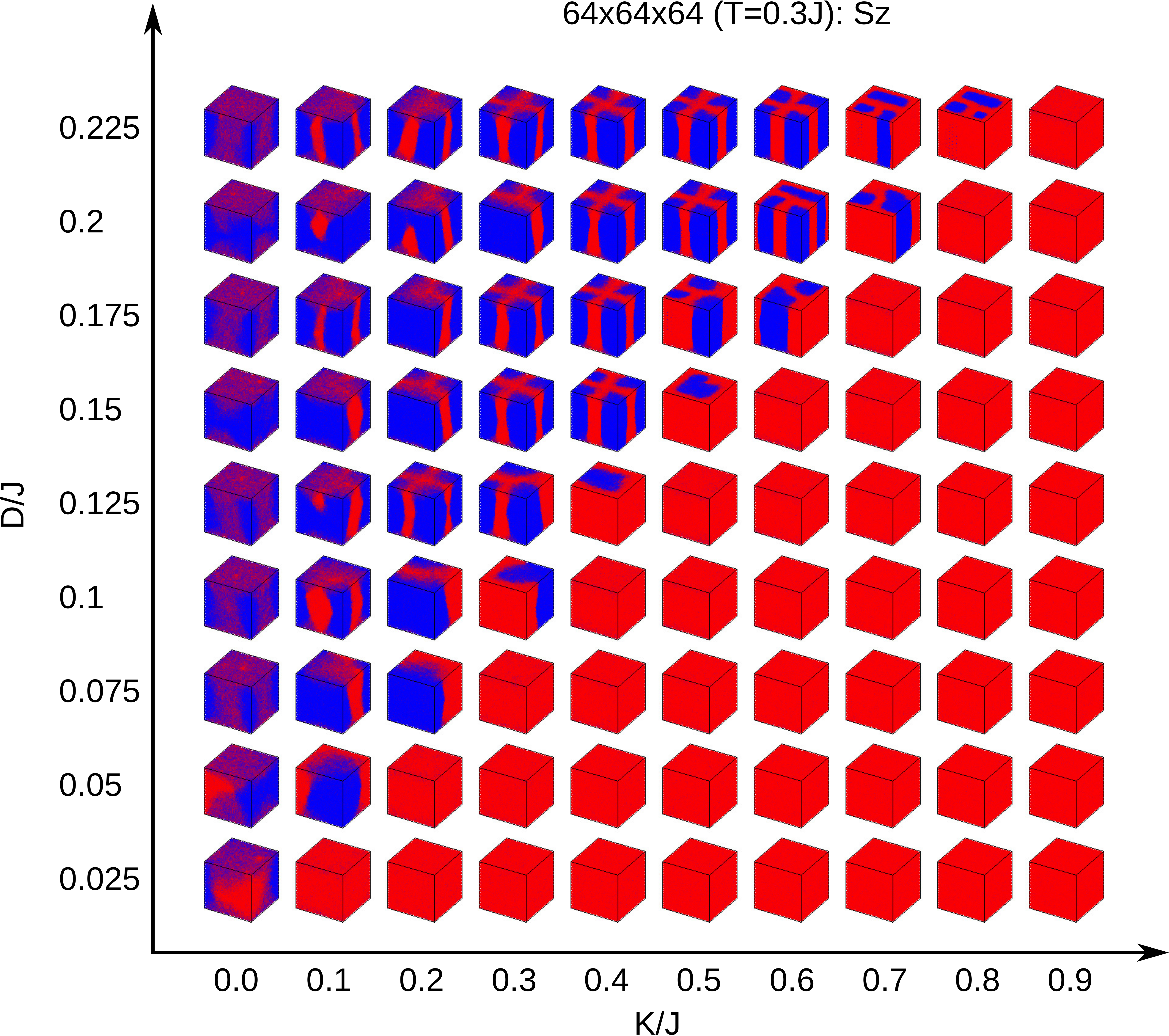}
\end{minipage}
\begin{minipage}{0.45\hsize}
\centering
\includegraphics[keepaspectratio,width=\hsize]{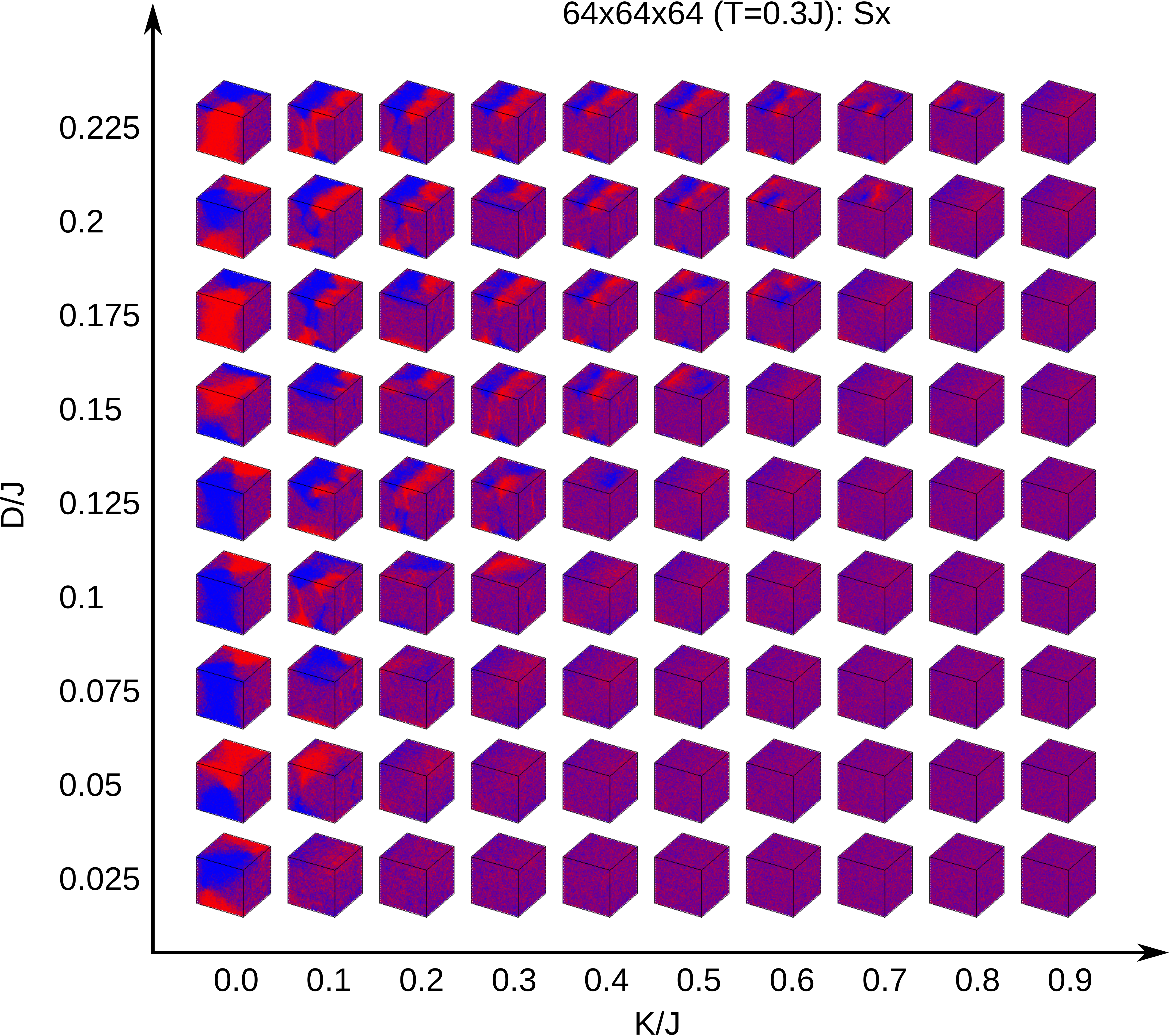}
\end{minipage}
\end{tabular}
\caption{Magnetic structures of $64\times64\times64$ systems under the field quench process from the out-of-plane ferromagnetic state with different anisotropies and DDIs.}
\label{fig:3d}
\end{figure}

\section{Conclusion and Discussion}\label{sec:conclusion}
In the present paper, we systematically surveyed magnetic configurations and presented a kind of diagram for $K/J$ and $D/J$ for multi-layered square-disk systems as a function of anisotropies, and DDI for various thicknesses. 
We found five distinct magnetic configurations, i.e., 
out-of-plane ferromagnetic state,
the in-plane ferromagnetic state,
the vortex state, the multi-domain state,
and the canted multi-domain state.
The vortex state and the canted multi-domain state appear specifically in systems with open boundary conditions.
Besides, we found that the canted multi-domain state, in which the $z$ component exhibits a concentric domain pattern, reflects the vortex structure in the in-plane axis.

We also presented microscopic configurations of the Neel cap structure.
The domain wall clearly shows the Neel type domain wall on the top and bottom of the system, while the Bloch type domain wall in the middle layers.
This structure can reduce the leaking magnetic flux as has been schematically pointed out in the literature.

We also studied the metastability of the out-of-plane ferromagnetic state by comparing the configurations obtained by the thermal-quench and field-quench processes. 
In some parameter regions, the field-quench process (Fig.~\ref{fig:ferrostart}) gave the out-of-plane ferromagnetic state 
while the thermal-quench process (Fig.~\ref{fig:eqstat}) gave a multi-domain state.
This difference indicates the metastability of the out-of-plane ferromagnetic state in the multi-domain region in the thermal-quench case. 
This metastability gives a mechanism of coercivity of relatively large grains in which DDI causes the multi-domain structure in the thermal-demagnetization process. 
We found that the collapse of the metastable state starts from a middle part of the system in contrast to nanosize systems, where the nucleation begins from a corner~\cite{Toga2020,Nishino2020}.
We also confirmed the scalability of magnetic configuration in different sizes of systems with the same aspect ratio, which indicates that the results of the present study are available for various sizes with the scaling despite the fact that DDI has a peculiar direction-dependent long-range nature of DDI. 

In the present paper, we mainly studied thin-film systems.
Recently developed experimental methods have made it possible to observe the magnetic structure in bulk~\cite{Suzuki2018}. 
As a primitive reference for the three-dimensional case, in Fig.~\ref{fig:3d}, we give magnetic configurations ($64\times 64\times 64$), in which we find a closed loop of magnetization reducing the stray field in small $K$ region.
In three-dimensional systems, the way to avoid the stray field is also three-dimensional.
Thus the vortex state around $K=0$ becomes a more complicated magnetic structure. 
Further studies for three-dimensional cases are left for future study, which will be essential for the coercivity of real magnets. 
\section*{Acknowledgements}
The authors would like to thank Dr. Satoshi Hirosawa for instructive discussion from experimental viewpoints, and to Professor Tetsuya Nakamura, Dr. Kentaro Toyoki, Dr. Motohiro Suzuki, and Dr. Yuta Toga for useful discussion on the result of the Monte Carlo study. The present work was supported by the Elements Strategy Initiative Center for Magnetic Materials (ESICMM) (Grant No. 12016013) funded by the Ministry of Education, Culture, Sports, Science and Technology (MEXT) of Japan, and was partially supported by Grants-in-Aid for Scientific Research C (No. 18K03444) from MEXT. The numerical calculations were performed on the Numerical Materials Simulator at the National Institute for Materials Science.

\appendix*
\section{Thickness dependence of DDI energies for a layered square system}
\begin{figure}[t]
	\includegraphics[keepaspectratio,width=0.8\hsize]{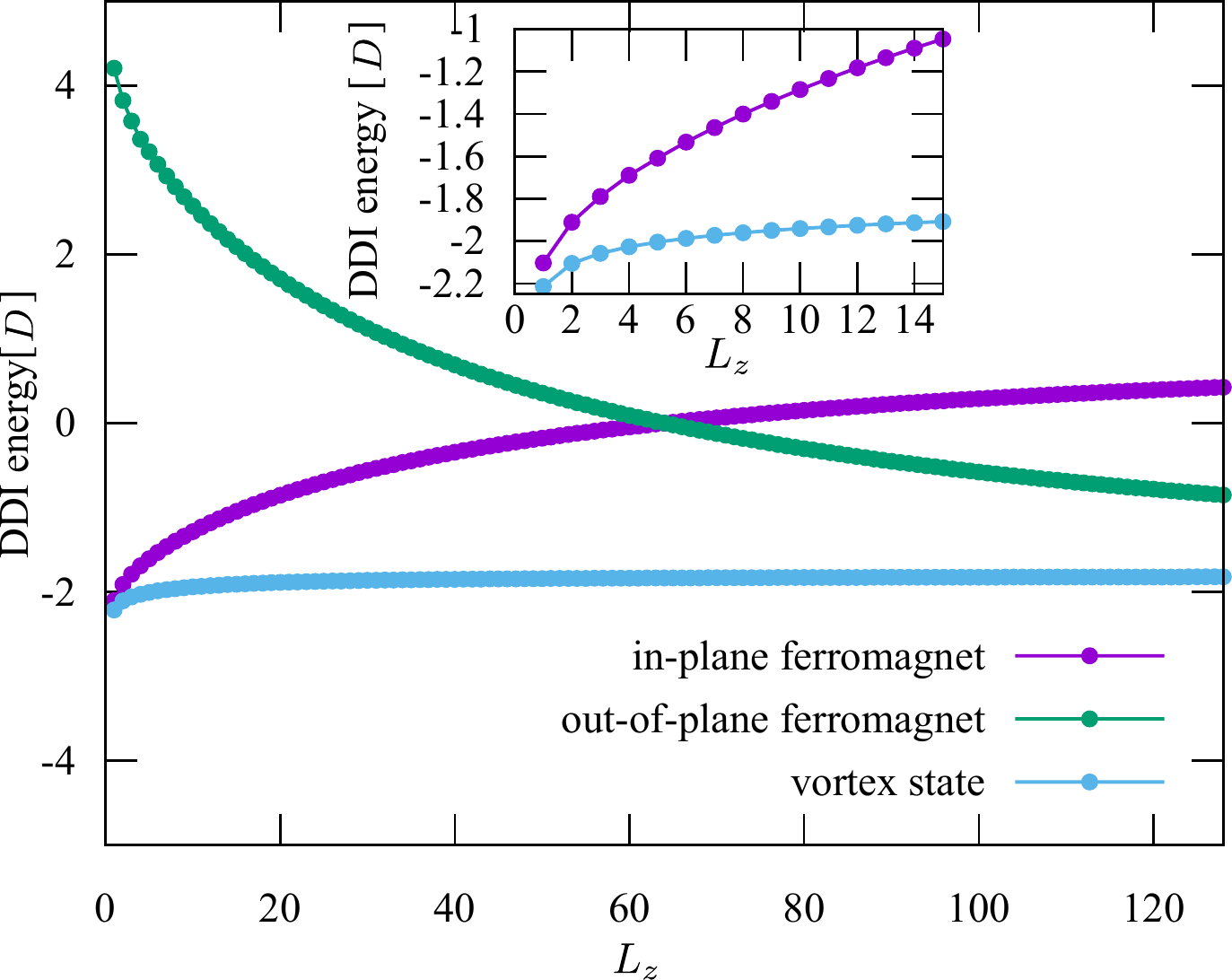}
\caption{
The total DDI energy per spin for forming the out-of-plane ferromagnetic state, the in-plane ferromagnetic state, and the vortex state as a function of $L_z$ for $64\times64\times L_z$ system.
}
\label{DDengLz}
\end{figure}

To study the effect of the thickness, we study the total energies of DDI for the out-of-plane ferromagnetic state, the in-plane ferromagnetic state, and the vortex state.
Figure~\ref{DDengLz} shows the DDI energy per spin for the $64\times64\times L_z$ system as a function of $L_z$.
Here, the magnetic configuration of each state is set as follows: all spins are aligned to the x-axis (the in-plane ferromagnetic state); all spins are aligned to the z-axis (the out-of-plane ferromagnetic state);
and spins at $i$ site is set as $[-r_{iy}+r_{cy},r_{ix},-r_{cx},0]/|\bm{r}_i-\bm{r}_c|$ (the vortex state), where $\bm{r}_c$ denotes the center of the system.

According to Fig.~\ref{DDengLz}, the vortex state has the lowest DDI energy in a whole range of $L_z$ in the present parameter range. 
In the in-plane ferromagnetic state, the DDI energy increases rapidly than the vortex state.
On the other hand, in the out-of-plane ferromagnetic state, the DDI energy decreases as the system thickens.
The DDI energy for the in-plane ferromagnetic state and that for the out-of-plane ferromagnetic state become the same value when the system is a cubic structure, $L_z=64$.
By using the result of Fig.~\ref{DDengLz}, we discuss the border of magnetic configurations in Sec.~\ref{sec:thermal}.

\bibliography{ref}
\end{document}